\documentclass[aps,prl,reprint,groupedaddress]{revtex4-1}

\usepackage{lipsum}
\usepackage{amsmath}	
\usepackage{amssymb}	
\usepackage{color} 		
\usepackage{graphicx} 	
\usepackage[T1]{fontenc}
\usepackage{times} 		
\usepackage{mathrsfs} 
\usepackage{myterms}
\usepackage{hyperref} 
\hypersetup{
    colorlinks=true,       
    linkcolor=blue,        
    citecolor=blue,        
    filecolor=magenta,     
    urlcolor=blue          
}
\usepackage[all]{hypcap} 

\newcommand{\dA}{\delta \hspace{-0.03cm} A}

\newcommand{\IR}{\text{\sc \fontfamily{ptm}\selectfont ir}}

\begin{document}

\title{Dark Photon Dark Matter from a Network of Cosmic Strings}

\author{Andrew J. Long}
\affiliation{Leinweber Center for Theoretical Physics, University of Michigan, Ann Arbor, Michigan 48109, USA}

\author{Lian-Tao Wang}
\affiliation{Kavli Institute for Cosmological Physics and Enrico Fermi Institute, University of Chicago, Chicago, Illinois 60637, USA}

\date{\today}

\begin{abstract}
We study the production of ultralight dark photons from a network of near-global, Abelian-Higgs cosmic strings.  We find that dark photons produced in this way are nonrelativistic today and can make up all of the dark matter for dark photon masses as small as $m_A \sim 10^{-22} \, \mathrm{eV}$.  
\end{abstract}


\maketitle

\section{\label{sec:intro}Introduction}

Although the evidence for a dark form of matter in the Universe is overwhelming, we still know next to nothing about its properties and interactions.  
In this article we will suppose that the dark matter is collection of nonrelativistic, elementary particles, and that the dark matter particle is bosonic with its spin equal to $1$ and its mass falling below approximately $1 \eV$.  
Such dark matter candidates have been called dark or hidden photons.  

There is a prolific and diverse literature on strategies for the detection of dark photon dark matter.  
Several notable techniques include the use of resonant cavities~\cite{Jaeckel:2007ch,Wagner:2010mi}, resonant LC circuits~\cite{Chaudhuri:2014dla}, accelerometers~\cite{Graham:2015ifn}, spin precession~\cite{Graham:2017ivz}, interferometers~\cite{Pierce:2018xmy}, periodic dielectric materials~\cite{Baryakhtar:2018doz}, observations of the $21 \cm$ radiation~\cite{Kovetz:2018zes}, observations of astrophysical heating~\cite{Bhoonah:2018gjb}, and gravitational superradiance~\cite{Arvanitaki:2010sy}.  
While the multitude of experimental probes is encouraging, there is also a sense in which these phenomenological studies are outpacing the theory work of model building.  

Models of dark photon dark matter suffer from a notorious production problem.  
In contrast with many familiar models of light \emph{scalar} dark matter (including the QCD axion and axionlike particles), light vector dark matter cannot be generated from the misalignment mechanism if it is minimally coupled to gravity~\cite{Dimopoulos:2006ms,Nelson:2011sf,Arias:2012az}.  
The misalignment energy density redshifts like $\rho \propto a^{-2}$ during inflation and dilutes by a factor of at least $e^{-120}$ if inflation lasts for more than $60$ $e$-foldings.  
Alternatively, dark photon dark matter can arise from inflationary quantum fluctuations (gravitational particle production), since the longitudinal polarization mode is not conformally coupled to gravity.  
Producing the observed dark matter relic abundance in this way requires a dark photon mass of $m_A \approx (10^{-5} \eV) (H_\mathrm{inf} / 10^{14} \GeV)^{-4}$~\cite{Graham:2015rva}, but since the inflationary Hubble scale is constrained to be $H_\mathrm{inf} \lesssim 10^{14} \GeV$, this mechanism is inadequate for smaller dark photon masses.  

Another approach to the dark photon production problem involves first populating an auxiliary sector, and then transferring energy to the dark photon.  
This strategy has been explored in several recent papers, which study the energy transfer from a scalar condensate into the dark photon via parametric resonance or tachyonic instability~\cite{Co:2018lka,Agrawal:2018vin,Bastero-Gil:2018uel,Dror:2018pdh}.  
Since the scale of the auxiliary sector is free to slide (within limits), one finds viable models of dark photon dark matter for a wide range of dark photon masses.  

This article discusses the production of dark photon dark matter from a network of cosmic strings.  
In the context of the preceding discussion, the cosmic string network serves as the ``auxiliary sector,'' which gradually transfers its energy into producing dark photons.  
One appealing feature of our scenario is that the cosmic strings follow from the same physics that gives rise to the massive dark photon.  
For instance, if the mass arises from a spontaneously broken local symmetry, then topology of the vacuum manifold implies the existence of a cosmic string solution, and causality arguments require a network of such strings to be formed in the Universe if symmetry breaking takes place after inflation is completed.  
Thus we would argue that cosmic strings provide a natural candidate for the source of dark photon dark matter.  
Earlier work on (nonaxion) dark matter production from defect networks can be found in Refs.~\cite{Jeannerot:1999yn,Lin:2000qq,Matsuda:2005fb,Cui:2008bd}; these studies did not consider the production of dark photon dark matter, which we will see, is more similar to the production of axion dark matter.  

The remainder of this article is organized as follows.  
In \sref{sec:model} we present a simple model for a massive dark photon, we detail this model's near-global cosmic string solution, and we discuss alternative models.  
In \sref{sec:production} we calculate the relic abundance of dark photons that arises from the evolution of the near-global, Abelian-Higgs string network in the scaling regime.  
We summarize our main results in \sref{sec:conc}.  

\section{\label{sec:model}Modeling dark photon-string coupling}

\subsection{\label{sub:AH_model}An Abelian-Higgs model}

Let $\Phi(x)$ be a complex scalar field and let $A_\mu(x)$ be a gauge potential vector field.  
Consider the Abelian-Higgs model, 
\begin{align}\label{eq:L}
	\Lscr = \bigl| D_\mu \Phi \bigr|^2 - \frac{1}{4} F_{\mu\nu} F^{\mu\nu} - \lambda \bigl( |\Phi|^2 - v^2 / 2 \bigr)^2 
\end{align}
where $D_\mu \Phi = \partial_\mu \Phi + i e A_\mu \Phi$ and $F_{\mu\nu} = \partial_\mu A_\nu - \partial_\nu A_\mu$.  
This theory has a $\U{1}$ gauge symmetry that is spontaneously broken by the scalar's vacuum expectation value, $\langle \Phi \rangle = v / \sqrt{2}$.  
The low-energy spectrum contains a vector particle, $A$, and a scalar singlet particle, $\rho$, with masses
\begin{align}
	m_A = e v 
	\qquad \text{and} \qquad 
	m_\rho = \sqrt{2 \lambda} v
	\per
\end{align}
As we will see shortly, we are interested in the near-global regime of this gauge theory, meaning
\begin{align}\label{eq:near_global}
	e^2 \ll 2 \lambda
	\qquad \text{and} \qquad m_A \ll m_\rho
	\com
\end{align}
which allows for efficient dark photon radiation from strings.  

\begin{figure}[t]
\hspace{0pt}
\vspace{-0in}
\begin{center}
\includegraphics[width=0.45\textwidth]{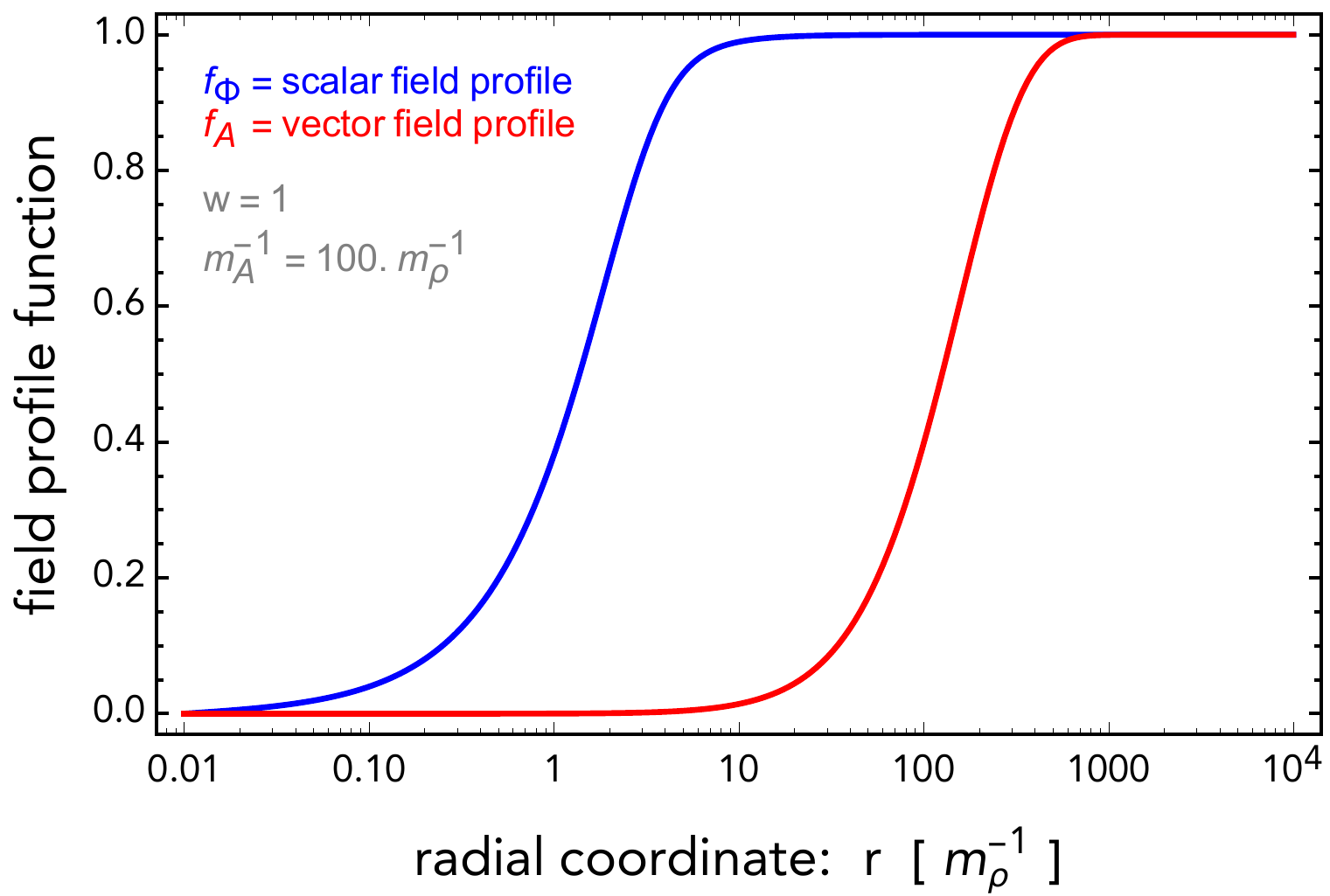}
\caption{
\label{fig:profiles}
The profile functions for a near-global, Abelian-Higgs cosmic string solution with winding number $w=1$ and $m_A^{-1} = 100 \, m_\rho^{-1}$.  
}
\end{center}
\end{figure}

This theory's classical field equations have an class of topological defect solutions known as Abelian-Higgs (AH) cosmic strings~\cite{Nielsen:1973cs} (for a review see \rref{VilenkinShellard:1994}).  
Strings are classified by their winding number, $w \in \Zbb$, and we are primarily interested in $w = \pm 1$.  
In cylindrical coordinates $\{ t, r, \varphi, z \}$, a long and straight string solution can be written as 
\begin{subequations}\label{eq:Ansatz}
\begin{align}
	\Phi & = \frac{v}{\sqrt{2}} \, \mathrm{f}_\Phi(r) \, e^{-i w \varphi} \\ 
	A_\mu & = \frac{w}{e} \, \mathrm{f}_A(r) \, \partial_\mu \varphi 
\end{align}
\end{subequations}
where the profile functions, $\mathrm{f}_\Phi(r)$ and $\mathrm{f}_A(r)$, satisfy the field equations for $\Phi$ and $A_\mu$.  
We have solved these equations numerically, and we present the results in \fref{fig:profiles}.  
The structure of the near-global, AH string generally consists of a narrow core at $0 \leq r \lesssim m_\rho^{-1}$ where the scalar field is displaced from the minimum of its potential and a wide cloud at $m_\rho^{-1} \lesssim r \lesssim m_A^{-1}$ where there is a nonzero magnetic flux.  

The string's tension (energy per length) arises primarily from a spatial gradient in the phase of $\Phi$, which remains nonzero well outside of the string core.  
Consequently the tension is logarithmically sensitive to large-distance-scale (IR) physics~\cite{Vilenkin:1982ks}.  
We can write the tension of a near-global Abelian-Higgs string with winding number $w$ as~\cite{VilenkinShellard:1994} 
\begin{align}\label{eq:mu}
	\mu(t) \approx \mu_0 \log[ m_\rho / m_\IR(t) ]
	\qquad \text{with} \qquad 
	\mu_0 \equiv \pi w^2 v^2 
	\com
\end{align}
where $m_\rho^{-1}$ is the length scale of the string core.  
The large-distance length scale, $m_\IR^{-1}$, corresponds to either the scale $m_A^{-1}$, beyond which $A_\mu$ compensates the gradient in $\Phi$, or the scale $d_\mathrm{sep}(t)$, giving the typical distance between neighboring strings at time $t$, and in general we can write $m_\IR^{-1} = \mathrm{min}[ m_A^{-1} , \, d_\mathrm{sep} ]$.  
We will see in the next section that $d_\mathrm{sep} \approx d_H / \sqrt{\xi}$ where $d_H(t)$ is the Hubble radius at time $t$ and $\xi(t)$ is the typical number of strings per Hubble volume.  
During the period of dark photon production we have $m_\IR(t) = H \sqrt{\xi}$.  

\subsection{\label{sub:coupling}Coupling of dark photons to the string}

To calculate the production rate of dark photons, we must first identify the coupling between this particle and the string.  
Let us consider perturbations around the string ansatz by substituting $\Phi \to \Phi \, e^{i \theta}$ and $A_\mu \to A_\mu + \dA_\mu$ where $\theta(x)$ is the would-be Goldstone boson field.  
The interactions of these field perturbations can be read off of the Lagrangian, 
\begin{align}\label{eq:L_int}
	\Lscr \supset 
	- \frac{1}{2} v^2 \bigl( 1 - \mathrm{f}_\Phi^2 \bigr) \bigl[ - w \bigl( 1 - \mathrm{f}_A \bigr) \, \partial \varphi + \partial \theta + e \, \dA \bigr]^2 
	\per
\end{align}
Recall that $1-\mathrm{f}_\Phi(r)$ only has support inside the string core.  
To study the interaction of relativistic dark photons with the string, the fields $\dA_\mu$ and $\partial_\mu \theta$ represent the transverse and longitudinal polarization states, respectively.  
The transversely polarized dark photon's interaction is suppressed by the small coupling, and the radiation of this polarization state is proportional to $e^2 \sim m_A^2 / v^2 \sim m_A^2 / \mu \ll 1$.  
On the other hand, \emph{longitudinally-polarized} dark photons couple to the string without any suppression from the small parameter $e$, or one can say that the $m_A^2$ suppression is compensated by a factor of $(E/m_A)^2$ as required by the Goldstone Boson Equivalence Theorem~\cite{Peskin:1995}, since the radiated dark photon's energy is larger than its mass.  

Hence in summary, we can neglect the emission of transversely polarized dark photons and model the radiation of longitudinally-polarized dark photons with the emission of the corresponding Goldstone boson.  
Goldstone emission from global strings is a well-studied process~\cite{Davis:1985pt,Vilenkin:1986ku}, in part because of its implications for axion dark matter~\cite{Davis:1986xc,Hiramatsu:2010yu}.  
We will adopt a standard result from this literature, which is that a global string radiates its Goldstone boson with a power (energy per time) given by $P \sim \mu$ where the coefficient is an $O(1)$ constant that depends weakly on the string's shape; see \sref{sub:radiation} for further discussion.  

\subsection{\label{sub:WGC}Implications of the weak gravity conjecture}

The theory presented here is perfectly well defined and predictive.  
However, it has been argued~\cite{Vafa:2005ui} that one may run into problems when attempting to embed low-energy effective theories, such as this one, into a self-consistent quantum theory of gravity.  
The essence of the problem is expressed by the weak gravity conjecture (WGC)~\cite{ArkaniHamed:2006dz}, which (very roughly) proposes that gravity should be the weakest force and prohibits the $e \to 0$ limit \pref{eq:near_global}.  
More to the point, the arguments used to motivate the WGC imply that a low-energy effective theory with charged matter should be extended at high energies to contain a tower of charged states at the mass scale $m_\mathrm{tower} \sim e \Mpl$.  
Since the low-energy effective theory does not describe the tower of charged states, one says that the WGC imposes a high-energy (UV) cutoff at $\Lambda_\mathrm{UV} \sim m_\mathrm{tower}$.  

However, the theory that we are considering here does not contain charged matter in the low-energy effective theory \pref{eq:L}, since our theory is in the Higgs phase, and therefore the arguments used to motivate the WGC are not justified.  
In particular, the original black hole thought experiment argument~\cite{ArkaniHamed:2006dz} does not apply in the Higgs phase where charge leaks off the black hole, and consequently the black hole can evaporate without emitting charged particles~\cite{Adler:1978dp,Cheung:2014vva}.  
The question then is whether the WGC can be bolstered by other arguments that lead to constraints on models in the Higgs phase~\cite{Cheung:2014vva,Heidenreich:2017sim,Reece:2018zvv,Craig:2018yld,Craig:2018yvw}.  
For instance, if the Compton wavelength of the dark photon is large compared to the size of the black hole then evaporation occurs slowly, and one might imagine some variant of the black hole thought experiment that applies to this case as well.  

Suffice it to say, our Abelian-Higgs toy model for the ultralight dark photon may or may not admit a self-consistent embedding into a quantum theory of gravity.  
However, we do not view violation of the WGC as a major shortcoming.  

\subsection{\label{sub:other_models}Alternative models for coupling dark photons to strings}

It is tempting to decouple the mass of the dark photon from the scale of the string and thereby avoid working in the $e^2 \ll 1$ regime.  
In this section, we discuss several alternative models for a low-mass dark photon coupled to a high-tension string.  

One possibility is to suppose that the string is superconducting and carries a current that sources the dark photon.  
Let us consider bosonic superconductivity~\cite{Witten:1984eb} as a simple example.  
This scenario requires two pairs of scalar and vector fields:  $(\phi, V_\mu)$ correspond to $\U{1}_\mathrm{string}$, while $(\sigma, A_\mu)$ correspond to $\U{1}_\mathrm{dark}$.  
The scalar potential has the form
\begin{align}
	V(\phi, \sigma) 
	& = \mu_\phi^2 |\phi|^2 + \lambda_\phi |\phi|^4 + \mu_\sigma^2 |\sigma|^2 
	\\ & \quad 
	+ \lambda_\sigma |\phi|^4 + \lambda_{\phi \sigma} |\phi|^2 |\sigma|^2 
	\nonumber
	\com
\end{align}
and the parameters are chosen such that $\langle \phi \rangle = v_\phi \neq 0$ and $\langle \sigma \rangle = v_\sigma = 0$ far away from the string~\footnote{For $v_\sigma$ strictly vanishing the dark photon is massless, but we do not expect the superconducting properties of the string to be qualitatively changed if $0 < v_\sigma \ll v_\phi$ instead.}.  
However $\phi \to 0$ at the string core, and here $\sigma$ develops a condensate.  
Given the proper initial condition, this condensate can support a current, $I_\mathrm{dark} = 2 \int \! \ud x \, \ud y \, \mathrm{Re}[ \sigma^\ast \partial_z \sigma]$, which acts as a source to let the string radiate dark photons with a power $P_{\! A} \sim I_\mathrm{dark}^2$~\cite{Vilenkin:1986zz}.  

In order to produce dark photons from superconducting strings, we must first address how the string's current develops.  
For electromagnetic superconductivity, the current arises via Faraday's law of induction when the cosmic string passes through a magnetic field, such as that found in galactic structure~\cite{Vilenkin:1982ks}.  
If a similar mechanism is responsible for charging up our hidden-sector strings, then additional degrees of freedom are required to generate the dark photon magnetic field.  
Without specifying this additional physics explicitly, one cannot estimate $I_\mathrm{dark}$, but in general the current cannot be made too large, $I_\mathrm{dark} \lesssim e_\mathrm{dark} v_\sigma / \sqrt{\lambda_\sigma}$, because otherwise the current becomes unstable~\cite{Vilenkin:1982ks}.  
If $I_\mathrm{dark}$ is allowed to be as large as $\mu^{1/2}$, then the emission of dark photons from a network of superconducting strings could make up all of the dark matter.  

Another possibility is to distinguish $\U{1}_\mathrm{dark}$ from $\U{1}_\mathrm{string}$, where the former is spontaneously broken giving mass to the dark photon, $A_\mu$, and the latter is broken at a much higher scale, forming the string from a new vector field, $V_\mu$~\cite{Long:2014mxa}.  
The two vector fields will interact through a gauge-kinetic mixing ($\epsilon$ term below).  
The relevant interactions are 
\begin{align}
	\Lscr &\supset - \frac{1}{4} V_{\mu \nu} V^{\mu \nu}  - \frac{1}{4} A_{\mu \nu} A^{\mu \nu} + \frac{\epsilon}{2} V_{\mu \nu} A^{\mu \nu} - V_\mu J^{\mu}_\mathrm{string} 
	\nn & \qquad 
	- A_\mu J^\mu_\mathrm{dark} + \frac{1}{2} m_V^2 V_\mu V^\mu + \frac{1}{2} m_A^2 A_\mu A^\mu 
	\com
\end{align}
where $J_\mathrm{string}^\mu$ and $J_\mathrm{dark}^\mu$ are the current densities of the matter charged under $\U{1}_\mathrm{string}$ and $\U{1}_\mathrm{dark}$, respectively.  
There is a magnetic flux of the $\U{1}_\mathrm{string}$ gauge field at the string and $V_{\mu\nu} \neq 0$, which becomes a source for the $\U{1}_\mathrm{dark}$ gauge field through the gauge-kinetic mixing term.  
In the mass eigenstate basis, the gauge kinetic mixing operator is replaced by $\Lscr_\mathrm{int} \sim (\epsilon m_A / m_V) \, A_\mu J_\mathrm{string}^\mu$, and one expects that the dark photon radiation power will be suppressed by $\epsilon^2 m_A^2 / m_V^2 \ll 1$.  
However, radiation of the longitudinal mode might be enhanced by a factor of $E^2/m_A^2$, thanks to the Goldstone boson equivalence theorem, possibly leading to the radiation of dark photons with only an $O(\epsilon^2)$ suppression.  
If there is also a light radial mode, $h_\mathrm{dark}$, accompanying the Higgsing of the $\U{1}_\mathrm{dark}$, we have an additional interaction $\sim \epsilon v  h_\mathrm{dark} A_\mu V^{\mu} $, which could lead to radiation of the dark photon suppressed by the size of the kinetic mixing.  
This interesting scenario merits further study.  

Let us close this section by speculating briefly on two other strategies for avoiding $e^2 \ll 1$.  
Up till now we have been assuming that the dark photon arises from an Abelian gauge field theory.  
Such theories also predict a radial mode, the scalar singlet $\rho$, but this degree of freedom does not play any role in dark photon production.  
Thus it is interesting to do away with $\rho$ by supposing either that the dark photon arises from a Stueckelberg theory or that the string is a fundamental string.  
(It is worth mentioning that swampland arguments may still apply to the Stueckelberg theory~\cite{Reece:2018zvv}.)  
We will say no more about these alternative models in the rest of the article, but it would be interesting to explore their implications more carefully.  

\section{\label{sec:production}Production of dark photon dark matter}

\subsection{\label{sub:formation}Cosmological formation of the string network}

Inflation evacuates the Universe of matter, and subsequently particles are created from decay of the inflaton condensate during the epoch of reheating.  
In the near-global regime \pref{eq:near_global}, it is reasonable to expect that reheating produces equal amounts of $\Phi$ particles and antiparticles, but only a negligible abundance of $A$ particles~\footnote{For instance, if the dark sector \pref{eq:L} is coupled directly to the inflaton, represented by the real scalar field $\phi(x)$, then $\Phi$ interacts through a mass-dimension-4 operator, $\Lscr_\mathrm{int} = - \kappa \phi^2 \abs{\Phi}^2$, but $A_\mu$ only interacts through mass-dimension-5 operators, $\Lscr_\mathrm{int} = c_1 M^{-1} \phi \abs{D_\mu \Phi}^2 - c_2 M^{-1} \phi F_{\mu\nu} F^{\mu\nu}$, assuming that $\phi(x)$ is not charged under the dark $\U{1}$.  Alternatively if the dark sector couples directly to the Standard Model, then $\Phi$ interacts through the Higgs portal operator, $\Lscr_\mathrm{hp} = - \kappa H^\dagger H \abs{\Phi}^2$, while $A_\mu$ couples via the gauge-kinetic mixing operator, $\Lscr_\mathrm{gkm} = - \epsilon/2 \, B_{\mu\nu} F^{\mu\nu}$, which leads to interactions that are suppressed by $e \epsilon \ll 1$.}.  
The $\Phi$ particles and antiparticles thermalize through their self-interaction~\footnote{This assumption imposes only a weak lower bound on the scalar self-coupling, $\lambda \gtrsim \sqrt{8 \pi T_\mathrm{pt} / \Mpl}$, which is easily satisfied for $\lambda = O(1)$.} reaching a temperature $T_\mathrm{rh}$, but the $A$ particles are extremely weakly interacting in the near-global regime \pref{eq:near_global}, preventing the dark photon from being produced thermally (negligible freeze-in)~\footnote{In the near-global regime, $e^2 \ll 1$, the thermal production of dark photons is dominated by bremsstrahlung, $\Phi + \bar{\Phi} \to \Phi + \bar{\Phi} + A$.  The yield is parametrically $Y_A^\mathrm{(fi)} \sim \lambda^2 e^2 \Mpl / T_\mathrm{pt} \sim m_\rho^2 m_A^2 \Mpl / v^5$, which is insensitive to initial conditions, such as the reheating temperature, but instead controlled by the plasma temperature at the phase transition.  For the parameter regime of interest, $Y_A^\mathrm{(fi)} \lll 1$ is entirely negligible.  Moreover, dark photons produced thermally have an initial energy set by the temperature, and for the parameter regime of interest, $m_A < 0.1 \eV$, these particles remain relativistic at radiation-matter equality, contributing to the dark radiation rather than dark matter. }.  

We assume that the $\U{1}$ gauge symmetry was restored during reheating, which imposes a lower bound on the temperature of the $\Phi$ particles and antiparticles, roughly $T_\mathrm{rh} \gtrsim v$~\footnote{More accurately, we need the thermal mass correction to exceed the tachyonic mass: $m_{\Phi,\mathrm{thermal}}^2(T_\mathrm{rh}) > \lambda v^2$.  If the thermal mass arises primarily from self-interactions then $m_{\Phi,\mathrm{thermal}}^2 \sim \lambda T^2$, and the condition imposes $T_\mathrm{rh} \gtrsim v$ independent of $\lambda$.  Alternatively, if the thermal mass comes from another interaction, such as a Higgs portal coupling $\Lscr_\mathrm{hp} = - \kappa H^\dagger H \abs{\Phi}^2$, then the bound is $T_\mathrm{rh} \gtrsim \lambda v / \kappa$.}.  
Subsequently, as the Universe expanded and cooled, the symmetry was broken through a cosmological phase transition.  
After the phase transition the $\Phi$ particles and antiparticles become scalar singlet particles $\rho$ and longitudinally polarized dark photons $A_L$.  
Eventually $\rho$ decays to Standard Model particles or $A$'s~\footnote{In the near-global regime \pref{eq:near_global}, the decay $\rho \to AA$ is extremely slow, and for some of the parameter space in \fref{fig:Omega} it exceeds the age of the Universe today.  If $\rho$ is cosmologically long lived, its relic abundance from production at the phase transition will overclose the Universe, and the model is not viable.  Extending the model to include a Higgs portal coupling, $\Lscr_\mathrm{hp} = - \kappa H^\dagger H \Phi^\ast \Phi$, opens the decay $\rho \to HH$.  For $\kappa = O(1)$, this decay occurs soon after the phase transition, and the relic abundance of $\rho$ is safely depleted.}, and the remaining $A$'s are relativistic, contributing a negligible amount of dark radiation~\footnote{The dark photons produced during the phase transition have an initial energy $E \sim T_\mathrm{pt}$, which redshifts to $E \sim T_\mathrm{eq} \sim 0.1 \eV$ at radiation-matter equality.   Provided that $m_A \lesssim 0.1 \eV$, these dark photons behave like dark radiation rather than dark matter.  The ones produced from the decay of $\rho$ have an initial energy $E \sim m_\rho$, which is typically even larger, making them more relativistic.  }.  

The symmetry-breaking phase transition causes a network of cosmic strings to be formed, as required by causality arguments~\cite{Kibble:1976sj,Zurek:1985qw}.  
A cosmic string network consists of a dynamically evolving collection of long and short string loops.  
Consider a time $t$ during the radiation era when the Hubble radius was given by $d_H(t) = 1/H(t) = 2t$.  
Short string loops have a length $L < d_H(t)$, and all points on a given loop are in causal contact, whereas long string loops have $L > d_H(t)$ and only a segment of the loop crosses any given Hubble volume.  
Short string loops are formed when either $d_H(t)$ grows to overtake a long string loop that was just outside the horizon or when long strings intersect and reconnect forming new loops.

This string network evolves toward a scaling regime by exhausting excess energy into the radiation of gravitational waves and particles.  
A segment of string at rest with length $L$ carries an energy of $E = \mu L$.  
By radiating away energy, curved string segments tend to straighten out, and string loops tend to shrink and decay.  

\subsection{\label{sub:radiation}Radiation from near-global, Abelian-Higgs strings}

In the previous section, we discussed how a string network maintains scaling by exhausting energy into radiation.  
Our near-global, Abelian-Higgs strings have four radiation channels: gravitational waves, scalar singlets, transversely polarized dark photons, and longitudinally polarized dark photons.  
In the remainder of this section we will discuss each of these radiation channels in turn, but let us first summarize the main results.  
A curved string segment with curvature scale $L$, such as a loop with radius $R = L$, radiates into each of the four channels with a power (energy emitted per time) given by 
\begin{subequations}\label{eq:power}
\begin{align}
	P_\mathrm{gw} & \sim \mu \times G_{\! N} \mu \\ 
	P_\rho & \sim \mu \times \Theta(1 - m_\rho L) \\  
	P_{\! A_T} & \sim \mu \times (m_A^2/\mu) \times \Theta(1 - m_A L) \\ 
	P_{\! A_L} & \sim \mu \times \Theta(1 - m_A L) 
	\per
\end{align}
\end{subequations}
where the step function $\Theta(z) = 1$ for $z > 0$ and $0$ otherwise.  

Accelerated string segments induce a quadrupole moment that results in gravitational wave radiation.  
The radiation power is parametrically $P_\mathrm{gw} \sim G_{\! N} \mu^2$ where $G_{\! N}$ is Newton's constant, and the dimensionless factor $G_{\! N} \mu \ll 1$ quantifies the coupling of gravity to the string.  
The coefficient can differ for gauge and global strings~\cite{Caldwell:1996en}, but in general $P_\mathrm{gw} \ll \mu$.  

Unlike gravitational wave radiation, the emission of massive particles can be kinematically blocked.  
For instance the motion of a periodically oscillating string loop with length $L$ and period $\tau = L/2$ can be decomposed into harmonic functions with angular frequency $\omega_n = 2 \pi n / \tau$ and $n \in \Zbb^+$.  
In order to radiate a particle of mass $m$ from mode $n$, the kinematic condition $\omega_n > m$ must be satisfied~\cite{Damour:1996pv}.  
If the particle is light with respect to the loop, in the sense that $mL < 4\pi$, then all of the $n > 1$ modes can radiate.  
However, heavy particles with $4 \pi < mL$ can only be radiated from higher harmonics, which are typically absent from smooth strings~\footnote{In addition to the perturbative radiation described in the main text, there can also be a nonperturbative radiation of the string-forming fields when antiparallel string segments overlap and annihilate.  This annihilation occurs during reconnection events, cusp evaporation, and the final stage of loop decay.  However these events are rare, and they can be neglected for the purposes of calculating the loop dynamics and dark photon relic abundance.}.  
In general, one can calculate how $P$ depends on $mL$, but since our results are insensitive to the specific form of this function, we will approximate it by a step, $\Theta(1 - mL)$, as in \eref{eq:power}.  

For the scalar singlet $\rho$ we generally have $m_\rho L \gg 1$, since the thickness of the string core is approximately $m_\rho^{-1}$.  
Therefore the radiation of the scalar singlet is always dramatically suppressed.  

The situation is more subtle for the dark photon, because it is parametrically lighter than the symmetry-breaking scale in the near-global regime \pref{eq:near_global}.  
Recall that only sub-Hubble loops and curved string segments can radiate, since larger-scale features are not yet in causal contact, and thus we are only interested in $L < d_H(t) \sim H^{-1}$ at any time $t$.  
There is a special time $t = t_\ast$ at which 
\begin{align}\label{eq:t_ast}
	m_A = H(t_\ast) 
	\qquad \text{and} \qquad 
	T(t_\ast) = \frac{\sqrt{m_A \Mpl}}{(\pi^2 g_\ast(t_\ast) / 90)^{1/4}}
\end{align}
where $g_\ast$ is the effective number of relativistic species.  
At early times ($m_A < H$) the dark photon is light for all sub-Hubble loops, $m_A L < m_A / H < 1$, and the radiation of dark photons is not kinematically suppressed.  
However, at late times ($H < m_A$) the radiation of dark photons from Hubble-scale loops is kinematically suppressed, since $m_A L \sim m_A / H > 1$, and these loops decay instead by gravitational wave emission.  
In this way, the near-global, Abelian-Higgs string network behaves like a network of global strings at early times, when the particle radiation is efficient, and a network of gauge strings at late times, when the particle radiation is suppressed.  

Furthermore, we can distinguish the radiation of the different dark photon polarization states.  
As we have already discussed in \sref{sub:AH_model}, the transverse polarization state couples to the string with an interaction strength of $e^2 \sim m_A^2 / \mu \ll 1$, and therefore the radiation power goes as $P_{\! A_T} \sim \mu \, (m_A^2/\mu) \, \Theta(1 - m_A L)$.  
On the other hand the longitudinal polarization state has an $O(1)$ interaction strength with the string, and the corresponding radiation power is $P_{\! A_L} \sim \mu \, \Theta(1 - m_A L)$.  

In summary, the emission of the scalar singlet, $\rho$, and the transversely polarized dark photon, $A_T$, can be neglected.  
At early times, $m_A < H$, the radiation is dominated by emission of longitudinally polarized dark photons, which contribute to the dark matter.  
At late times, $H < m_A$, the gravitational wave radiation is dominant.  

\subsection{\label{sub:structure}Structure of the string network}

The efficiency with which individual loops radiate and decay affects the qualitative structure of the network as a whole.  
Consider a loop at rest of length $L(t)$ at time $t$ that radiates with a power $P(L,t)$.  
The energy carried by this loop is $E(t) = \mu L(t)$, and it decreases according to 
\begin{align}
	\frac{dE}{dt} = - P 
	\per
\end{align}
If $P$ is independent of both $L$ and $t$ then the solution is simply $L(t) = L_1 - (P/\mu) \, \bigl( t-t_1 \bigr)$.  
The loop has completely decayed away at time $t = t_L$ such that $L(t_L) = 0$, which corresponds to $t_L = t_1 + L_1 / (P/\mu)$.  
The number of elapsed Hubble times is given by $(t_L - t_1) / t_1 = (L_1 / t_1) \, (P/\mu)^{-1}$, and therefore an initially Hubble-scale loop, $L_1 \sim t_1$, decays in approximately $(P/\mu)^{-1}$ Hubble times.  

We learned in \sref{sub:radiation} that for near-global, Abelian-Higgs strings we should distinguish the early time behavior when $m_A < H(t)$ from the late time behavior when $H(t) < m_A$.  
At early times, Hubble-scale loops predominantly radiate longitudinally polarized dark photons and decay quickly, in $(P/\mu)^{-1} \sim (P_{\! A_L}/\mu)^{-1} \sim 1$ Hubble times.  
At late times, these loops mainly radiate gravitational waves and it takes many Hubble times for them to decay:  $(P/\mu)^{-1} \sim (P_\mathrm{gw}/\mu)^{-1} \sim (G_{\! N} \mu)^{-1} \gg 1$.  

In this way the network of near-global, Abelian-Higgs strings acts like a chimera; it resembles a network of global strings at early times and a network of gauge strings at late times.  
If we take a snapshot of a global string network, it will typically display several straight string segments crossing a given Hubble volume and perhaps one big loop.  
If there is a loop, it will radiate and decay in $O(1)$ Hubble times, and eventually a new loop will be formed from the intersection and reconnection of the long strings.  
However, a snapshot of a gauge string network will contain many loops, since they can only decay by the emission of gravitational waves.  
In this sense, our near-global, Abelian-Higgs string network experiences a phase transition at time $t = t_\ast$ when $m_A = H(t_\ast)$ from a long-string-dominated structure to a loop-dominated structure.  

\subsection{\label{sub:scaling}The network exhausts energy to maintain scaling}

To calculate the dark photon relic abundance in the next section, we must determine how much energy is being radiated by the string network into dark photons.  
We have done half of this calculation in \sref{sub:radiation} by estimating the dark photon radiation power from a given string segment.  
By taking the expressions for $P_{\! A_T}$ and $P_{\! A_L}$ from \eref{eq:power} and using what we have learned about the structure of the string network in \sref{sub:structure}, we could calculate the corresponding radiation power densities, $\Pcal_{\! A_T}$ and $\Pcal_{\! A_L}$, produced by the entire network.  
However, in the remainder of this section we will apply a less direct, but more adaptable line of argument.  

Let $\rho_\mathrm{str}(t)$ be the energy density of the string network at time $t$, coarse grained on a length scale that is large compared to the typical distance between neighboring strings.  
In \sref{sub:structure} we argued that most of the network's energy is carried by long strings at early times [$m_A < H(t_\ast)$].  
Therefore the energy density should obey~\footnote{If we could abruptly shut off the radiation ($\Pcal \to 0$), then the ``free'' network of long strings would obey $\rho_\mathrm{str} \propto a(t)^{-2}$, where $a(t)$ is the scale factor at time $t$, and $H = \dot{a}/a$.  }
\begin{align}\label{eq:dot_rho_str}
	\dot{\rho}_\mathrm{str} + 2 H \rho_\mathrm{str} = - \Pcal
	\quad \text{for} \quad t < t_\ast
	\per
\end{align}
The power density $\Pcal(t)$ is the total amount of energy radiated from the string network per unit time and volume at time $t$.  
We can define the dimensionless function $\xi(t)$ by writing 
\begin{align}\label{eq:rho_str}
	\rho_\mathrm{str}(t) = \xi(t) \, \mu(t) \, / \, t^2 
	\com
\end{align}
without any loss of generality.  
Since a Hubble-length segment of string contributes $\sim \mu d_H / d_H^3 \sim \mu / t^2$ to the energy density, we can interpret $\xi(t)$ as the average number of long strings per Hubble volume.  
Requiring \eref{eq:rho_str} to be a solution of \eref{eq:dot_rho_str} implies~\cite{Harari:1987ht} (see also \rrefs{Hagmann:2000ja,Gorghetto:2018myk})
\begin{align}\label{eq:Pcal}
	\Pcal
	= \frac{\xi \mu}{t^3} \left( 1 - \frac{t \dot{\mu}}{\mu} - \frac{t \dot{\xi}}{\xi} \right) 
	\approx 2 H \rho_\mathrm{str}
	\per
\end{align}
The radiation power density for longitudinally polarized dark photons is now simply $\Pcal_{\! A_L} \approx \Pcal \, \Theta(t - t_\ast)$, since we argued in \sref{sub:radiation} that this radiation channel dominates at early times.  

In the preceding discussion it seems that we have simply traded the unknown function $\Pcal_{\! A_L}(t)$ for the unknown function $\xi(t)$.  
Moreover the logic of this relation is apparently backward, since it is the radiation of dark photons that determines the structure and energy density of the string network, parametrized by $\xi$, and not $\xi$ that controls $\Pcal_{\! A_L}$.  
However, it is convenient to parametrize our ignorance in terms of $\xi(t)$, since this quantity can be measured with numerical simulations.  

In fact the time dependence of $\xi(t)$ has attracted significant attention lately.  
For a string network in the scaling regime, $\xi$ should be a constant~\cite{VilenkinShellard:1994}.  
This expectation is confirmed by various numerical simulations of Abelian-Higgs (gauge) strings~\cite{Hindmarsh:2017qff,Correia:2018gew} Nambu-Goto strings~~\cite{Blanco-Pillado:2013qja}, and global strings~\cite{Moore:2001px,Lopez-Eiguren:2017dmc}.  
However, several recent numerical simulations~\cite{Yamaguchi:1998iv,Klaer:2017qhr,Gorghetto:2018myk,Kawasaki:2018bzv} and analytical arguments~\cite{Martins:2018dqg} have begun to indicate that a network of \emph{global strings}, such as axion strings, may exhibit a logarithmic deviation from scaling.  
In our notation, the results of \rref{Gorghetto:2018myk} are summarized as 
\begin{align}\label{eq:xi}
	\xi(t) \ \simeq \ \xi_0 \log [ m_\rho / H ]
	\qquad \text{with} \qquad 
	\xi_0 \simeq 0.2 
	\per
\end{align}
The logarithm appearing here is in addition to the well-known one that comes from $\mu(t)$; see \eref{eq:mu}.  
A logarithmic deviation from scaling would have important implications for models of axion dark matter~\cite{Yamaguchi:1998iv,Gorghetto:2018myk,Kawasaki:2018bzv,Martins:2018dqg}.  However, it is important to bear in mind that the alleged deviation from scaling is still a matter of active research, and that evidence supporting this conclusion primarily comes from string network simulations, which are inherently limited by their dynamic range.  In order to quantify the effect of these uncertainties on the production of dark photons, we will calculate the relic abundance using both \eref{eq:xi} as well as taking simply $\xi(t) = \xi_0 = 2$ as suggested by Refs.~\cite{Moore:2001px,Lopez-Eiguren:2017dmc}.  

Simulations of near-global, Abelian-Higgs string networks in our parameter regime of interest are not available to provide an estimate of $\xi(t)$.  
Nevertheless, the arguments in \sref{sub:structure} imply that the structure and evolution of this network at early times [$m_A < H(t_\ast)$] should be similar to that of a global string network.  
Therefore we are motivated to adopt the results of the recent global string network simulations.  
Using the expressions for $\mu(t)$ and $\xi(t)$ from \erefs{eq:mu}{eq:xi} we evaluate $\Pcal_{\! A_L} \approx \Pcal$ from \eref{eq:Pcal} to obtain 
\begin{align}\label{eq:Pcal_AL}
	\Pcal_{\! A_L} \approx \frac{\xi(t) \mu(t)}{t^3} \, \Theta(t_\ast - t)
\end{align}
where $H(t) = 1/(2t)$ during the radiation era.  
The logarithmic factors can be quite large:  $\log m_\rho / H(t_\ast) \simeq 20$ for $m_\rho = 10^{15} \GeV$ and $H(t_\ast) = m_A = 10^{-22} \eV$.  
Note that \eref{eq:Pcal_AL} is only valid at early times, corresponding to $t < t_\ast$ or equivalently $m_A < H(t)$, while the near-global, Abelian Higgs string network behaves like a global string network.  
At later times, the expression for $\xi$ from \eref{eq:xi} is no longer applicable, but moreover the emission of dark photons from Hubble-scale loops is kinematically suppressed, so we simply set $\Pcal_{\! A_L} = 0$ for $t > t_\ast$.  

\subsection{\label{sub:relic}Relic abundance of radiated dark photons}

Let $\rho_{A}(t)$ denote the energy density of dark photons at time $t$.  
Since most dark photons are relativistic at early times, $m_A < H(t)$, their energy density satisfies
\begin{align}\label{eq:dot_rho_A}
	\dot{\rho}_{A} + 4 H \rho_{A} = \Pcal_{\! A_T} + \Pcal_{\! A_L} 
\end{align}
where $\Pcal_{\! A_T}$ and $\Pcal_{\! A_L}$ represent the emission of transversely and longitudinally polarized dark photons from the strings.  
The arguments in \sref{sub:radiation} imply that $\Pcal_{\! A_T}$ is negligible in the near-global regime, and the calculations in \sref{sub:scaling} give $\Pcal_{\! A_L}$ in \eref{eq:Pcal_AL}.  
The solution of \eref{eq:dot_rho_A} is given by~\footnote{We have closely followed the work of \rref{Gorghetto:2018myk}, which derives this expression for axion dark matter.  }
\begin{align}\label{eq:rho_A}
	\rho_{A}(t) & 
	\approx \frac{4}{3} \xi \mu H^2 \log m_\rho / H
	\qquad \text{for} \qquad 
	t < t_\ast
\end{align}
up to terms that are suppressed by powers of $\log m_\rho / H$.  
Due to the logarithmic growth in $\mu$ and $\xi$, the integral is dominated by late times.  
If we had not taken the time dependence of $\mu$ and $\xi$ into account, then we would have obtained $\rho_{A} = (\xi \mu / t^2) \log t/t_\mathrm{initial}$ instead, and each logarithmic time interval would have contributed equally.  

The dark photons radiated at early times will be relativistic, but they will eventually become nonrelativistic through redshifting.  
Thus, to determine the axion relic abundance, we should calculate the number density of dark photons, denoted by $n_A$.  
This density evolves according to 
\begin{align}\label{eq:dot_n_A}
	\dot{n}_A + 3 H n_A = \Scal_A
\end{align}
where the source density $\Scal_A(t)$ is the number of dark photons produced per unit time and volume at time $t$.  
The power density $\Pcal_A$ and source density $\Scal_A$ are related by 
\begin{align}\label{eq:Scal_A_spectrum}
	\Scal_A = \int_0^\infty \! \! \ud k \, \frac{d\Scal_A}{dk} = \int_0^\infty \! \! \ud k \, \frac{1}{E_k} \frac{d\Pcal_A}{dk}
\end{align}
where we integrate over the magnitude of the momentum of the radiated particles, and $E_k = \sqrt{k^2 + m_A^2}$ is their corresponding energy.  

In general, we expect the spectrum $d\Pcal_A/dk$ to peak at an energy $E_k = O(1) \times H$.  
This is because the emission of dark photons from near-global, Abelian-Higgs strings should be similar to the emission of Goldstone bosons from global strings.  
Since the radiation is very efficient, as we discussed in \sref{sub:radiation}, we expect that as soon as a curved string segment enters the horizon it will quickly radiate away its excess energy and straighten out or that a loop will quickly collapse.  
This leads to a peaked spectrum at an energy $E_k = O(1) \times H$.  
However, there is a large uncertainty associated with the spectral index; if the spectrum has a high-momentum tail, then each particle carries more energy, and there are fewer particles radiated.  
Typically a direct calculation of the spectrum relies on numerical simulations of the string network, e.g., \rrefs{Yamaguchi:1999yp,Hiramatsu:2010yu,Gorghetto:2018myk}, which is beyond the scope of our work.  
We take a simplified approach by assuming that the spectrum of particles radiated at time $t$ is monochromatic with energy $E_k = \bar{E}_A(t) \sim H(t)$ where $x_0 = O(1)$.  
This lets us write the source density as 
\begin{align}\label{eq:Scal_A}
	\Scal_A \approx \frac{1}{\bar{E}_A} \Pcal_A 
	\qquad \text{with} \qquad 
	\bar{E}_a(t) \sim H(t)
	\per
\end{align}
It would be straightforward to extend this analysis to cover a power law spectrum with variable spectral index, but we leave that generalization to future work.  

With the preceding assumptions and caveats about the spectrum, we now proceed to solve for the density of dark photons.  
The solution of \eref{eq:dot_n_A} is given by~\footnote{We have followed closely the work of \rref{Gorghetto:2018myk}, which derives this expression for axion dark matter.  } 
\begin{align}\label{eq:n_A}
	n_A(t) 
	\approx \frac{8}{\bar{E}_A/H} \xi \mu H 
	\qquad \text{for} \qquad 
	t < t_\ast
\end{align}
where we have dropped subdominant terms.  
Note that the total density of dark photons is decreasing with time due to the cosmological dilution, $n_A \sim t^{-1}$, but the comoving density is growing, $a^3 n_A \propto t^{3/2} n_A \propto t^{1/2}$.  
The relic abundance of dark photons today (time $t=t_0$) is given by 
\begin{align}\label{eq:Oh2}
	\Omega_A h^2 
	= \frac{m_A \, Y_A(t_0) \, s(t_0)}{3 H_0^2 \Mpl^2 / h^2}
\end{align}
where $H_0 = 100 h \km / {\rm sec} / {\rm Mpc}$ is the Hubble constant and $\Mpl \simeq 2.43 \times 10^{18} \GeV$ is the reduced Planck mass.  
Here, we have also introduced the yield, $Y_A(t) = n_A(t) / s(t)$, where $s = (2\pi^2/45) g_{\ast S}(t) T(t)^3$ is the cosmological entropy density at time $t$ when the plasma temperature is $T(t)$.  
Dark photon radiation becomes negligible at $t = t_\ast$, and afterward the yield is conserved, $Y(t_0) = Y(t_\ast)$.  
Then, using the expression for $n_A(t_\ast)$ from \eref{eq:n_A}, we have 
\begin{align}\label{eq:Oh2_final}
	\Omega_A h^2 
	& \simeq 
	\bigl( 0.12 \bigr) 
	\left( \frac{m_A}{10^{-13} \eV} \right)^{1/2} 
	\left( \frac{\sqrt{\mu(t_\ast)}}{10^{14} \GeV} \right)^{2} 
	\\ & \qquad \times 
	\left( \frac{\xi(t_\ast)}{16} \right) 
	\left( \frac{\bar{E}_A}{H} \right)^{-1} 
	\left( \frac{H(t_\ast)}{m_A} \right)^{-1/2} 
	\, , \nonumber 
\end{align}
where we have taken the effective number of relativistic species to be $g_\ast = g_{\ast S} = 106.75$.  

\section{\label{sec:conc}Discussion and Conclusions}

\begin{figure}[t]
\hspace{0pt}
\vspace{-0in}
\begin{center}
\includegraphics[width=0.40\textwidth]{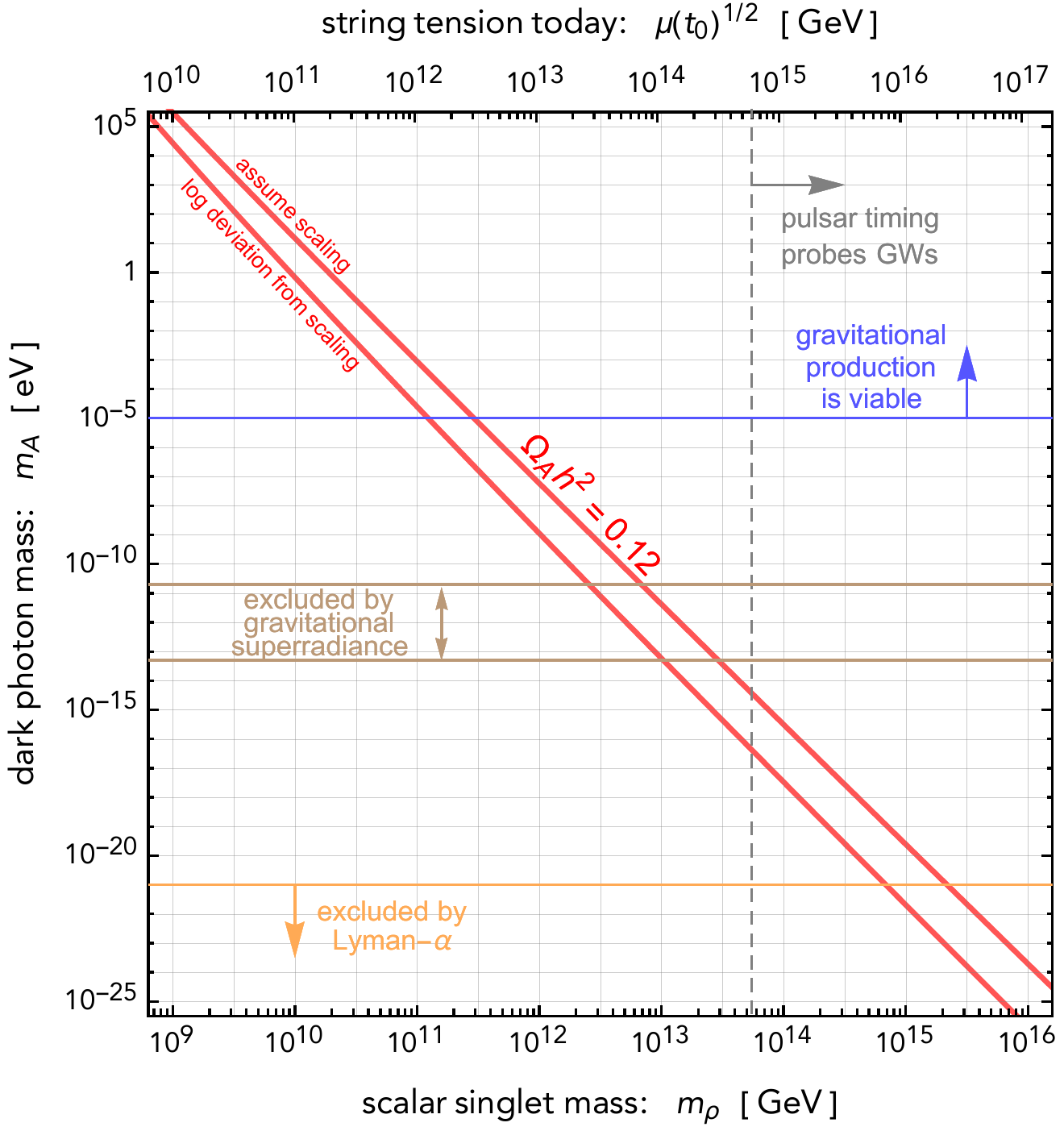}
\caption{
\label{fig:Omega}
The relic abundance of dark photon dark matter, given by \eref{eq:Oh2_final}, matches the observed dark matter relic abundance along the red lines labeled ``$\Omega_A h^2 \simeq 0.12$'' for an interesting region of parameter space where the dark photon's mass is sub-eV and the scale of symmetry breaking is somewhat below the GUT scale.  The two red lines serve to quantify the uncertainty in our calculation associated with evolution of the string network.  
}
\end{center}
\end{figure}

We show the relevant parameter space in \fref{fig:Omega}.  
Since the model has four free parameters $(v, \lambda, e, T_\mathrm{rh})$, we show only the two-dimensional slice of parameter space with $\lambda = 1$.  
Our results are insensitive to the postinflationary reheat temperature, $T_\mathrm{rh}$, as long as it is high enough for symmetry restoration; see the discussion in \sref{sub:formation}.  
The value of the string tension today is given by \eref{eq:mu}, which evaluates to $\mu(t_0) \approx (\pi/2\lambda) \, m_\rho^2 \, \log[m_\rho / m_A]$, and since this is only logarithmically sensitive to the dark photon mass, we fix $m_A = 10^{-10} \eV$ and show the corresponding value of $\mu(t_0)$ on the top of the plot.  

Recall from the discussion in the Introduction that the problem of dark photon dark matter production can be solved by inflationary quantum fluctuations (gravitational particle production) for $m_A \gtrsim 10^{-5} \eV$~\cite{Graham:2015rva}; this is indicated by the blue line in \fref{fig:Omega}.  
Additionally, models of particle dark matter with mass $m \lesssim 10^{-21} \eV$ are inconsistent with probes of cosmological structure, namely Lyman-$\alpha$ forest observations~\cite{Hui:2016ltb}; this is indicated by the orange line in \fref{fig:Omega}.  

Along the diagonal red lines, the relic abundance of longitudinally polarized dark photons matches the measured dark matter relic abundance, $\Omega_\mathrm{dm} h^2 \simeq 0.12$.  
Larger values of $m_A$ and $m_\rho$ (above the red line) are ruled out, because dark photon dark matter is overproduced.  
Regarding the dark photon production problem that we discussed in the Introduction, it is clear from these results that dark photon dark matter can be produced from its own near-global, Abelian Higgs cosmic string network for a wide range of dark photon masses.  
Models with smaller dark photon masses allow for viable dark matter production as long as they have correspondingly higher symmetry breaking scales, represented here by the string tension and the scalar singlet mass.  

The symmetry breaking scale is bounded from above in two ways.  
In order to form the string network via a cosmological phase transition, the symmetry must be restored after inflation.  
This imposes a lower bound on the postinflationary reheating temperature, $T_\mathrm{rh}$.  
For the Abelian-Higgs model we have studied here, this bound is roughly $T_\mathrm{rh} \gtrsim v$; see the discussion in \sref{sub:formation}.  
On the other hand, measurements of the cosmic microwave background constrain the energy scale of inflation~\cite{Aghanim:2018eyx}, which implies an upper bound on the reheating temperature that is at least as strong as $T_\mathrm{rh} \lesssim 10^{16} \GeV$ and possibly stronger depending on the model of inflation and reheating.  
Taken together these constraints imply $v \lesssim 10^{16} \GeV$ or $\sqrt{\mu(t_0)} \lesssim v \log^{1/2} \sim 10^{17} \GeV$.  
Thus, we conclude that the parameter space shown in \fref{fig:Omega} can still be consistent with cosmological limits on the symmetry breaking scale.  

Gravitational wave radiation provides a more direct test of the symmetry breaking scale.  
As we have discussed in \sref{sub:radiation} the collapse of string loops produces gravitational wave radiation, which is expected to survive in the Universe today as a stochastic gravitational wave background~\cite{Caprini:2018mtu}.  
Pulsar timing array (PTA) observations provide stringent constraints on the presence of such a gravitational wave radiation in the Universe today.  
For a network of Nambu-Goto or Abelian-Higgs cosmic strings, the loops are long lived and their gravitational wave emission should be observed by PTA measurements if the string tension is high enough, leading to constraints at the level of $\mu^{1/2} \lesssim 6 \times 10^{14} \GeV$~\cite{Blanco-Pillado:2013qja} (see also \rref{Sanidas:2012ee}), which would naively rule out everything to the right of the gray dashed line in \fref{fig:Omega}.  
However, for a network of global strings, the loops decay quickly by Goldstone boson emission, and the predicted gravitational wave signal is not within reach of PTA limits, which leaves the string tension unconstrained.  
As we have discussed in \sref{sub:structure} our near-global, Abelian-Higgs string network behaves like a global string network at early times and like a conventional gauge string network at late times.  
This opens the possibility that the PTA limits on $\mu^{1/2}$ are relaxed for the near-global, Abelian-Higgs strings, but we leave a detailed investigation of this point for future work.  

In order to quantify the uncertainties in our calculation associated with the evolution of the string network, we have evaluated the dark photon relic abundance, assuming both that the string network reaches the scaling regime, corresponding to $\xi(t)  \simeq 2$ in \eref{eq:rho_str}, and that the network exhibits a logarithmic deviation from scaling, corresponding to $\xi(t) \propto \log t$ as in \eref{eq:xi}.  
The behavior of global string networks is currently a matter of active research and debate.  
Since the logarithm at $t_\ast$ can be as large as approximately $80$, the effect on dark photon production is to shift the favored region of parameter space by roughly one decade.

In the Introduction we have enumerated various experimental strategies for probing dark photon dark matter. 
Most of these tests rely on a direct (nongravitational) interaction between the dark sector and the Standard Model, which is absent from the model we presented in \sref{sub:AH_model}.  
However, the effect of gravitational superradiance requires only a minimal gravitational coupling, and the observations of rapidly-spinning black holes place constraints on ultralight dark photons (whether or not they are the dark matter)~\cite{Baryakhtar:2017ngi}.
It would be interesting and fruitful to explore extensions of our minimal toy model with additional interactions.  
Interactions mediated by the Higgs-portal coupling would not lead to detectable signatures in low-energy observables, since the heavy scalar singlet is off shell and its effects at energy $E \ll m_\rho$ are suppressed by powers of $(E/m_\rho)^2$.  
Additionally, it would be interesting to suppose that the gauged $\U{1}$ symmetry is associated with one of the nonanomalous global symmetries of the Standard Model: $\mathsf{B} - \mathsf{L}$, $\mathsf{L}_e - \mathsf{L}_\mu$, and $\mathsf{L}_\mu - \mathsf{L}_\tau$.  

Let us close by recalling that dark photon dark matter arising from a network of near-global, Abelian-Higgs cosmic strings is necessarily longitudinally polarized at the time of production.  
As we discussed in \sref{sub:radiation} this can be understood from the perspective of the Goldstone-boson equivalence theorem.  
It would be interesting to explore how the polarization is affected by cosmological structure formation and to investigate strategies for testing the polarization of dark photon dark matter.  

\begin{acknowledgments}
\paragraph*{Acknowledgements.}  
We thank Matt Reece and Grant Remmen for discussions of the weak gravity conjecture.  
A.J.L. \ is grateful to Aaron Pierce and Tanmay Vachaspati for discussions of Goldstone boson radiation from topological defects.  
A.J.L. and L.T.W. \ are supported by the U.S. Department of Energy under Grants No. DE-SC0007859 and No. DE-SC0013642, respectively. 
\end{acknowledgments}

\bibliography{dark_photon}

\end{document}